\documentclass{article}

\usepackage[dvips]{graphicx}

\usepackage{psfrag}

\usepackage{latexsym,amssymb}

\newcommand{\mylab}[3]{\raisebox{#2}[0mm][0mm]{%
\makebox[0mm][l]{\hspace*{#1}\textbf{#3}}}}
\def\spacce#1{\hskip #1pt}
\def\drawline#1#2{\raise 2.5pt\vbox{\hrule width #1pt height #2pt}}
\def\solid{\drawline{24}{.5}\nobreak\ }

\def\bdash{\hbox{\drawline{3}{.5}\spacce{2}}}

\def\dashed{\bdash\bdash\bdash\bdash\nobreak\ }

\def\bdot{\hbox{\drawline{1}{.5}\spacce{2}}}

\def\dotted{\hbox{\leaders\bdot\hskip 24pt}\nobreak\ }

\def\circle{$\circ$\nobreak }
\def\trian{\raise 1.25pt\hbox{$\scriptstyle\triangle$}\nobreak\ }
\def\dtrian{\raise 1.25pt\hbox%
{$\scriptscriptstyle\bigtriangledown$}\nobreak\ }
\def\cuadrao{\raise 1.25pt\hbox{$\scriptstyle\Box$}\nobreak\ }
\def\diamon{\raise 1.25pt\hbox{$\scriptstyle\diamond$}\nobreak\ }

\def\squar{\raise 1.25pt\hbox{$\scriptstyle\Box$}\nobreak}
\def\AJ02{del \'Alamo \& Jim\'enez 2002}

\def\HUSS{Hussain \& Clark, 1981}
\def\Wills{Wills, 1964}
\def\alajim01{del \'Alamo \& Jim\'enez, 2001}
\def\Risoe{Nielsen {\it et al.}, 2002}

\def\Harmoni{Olesen, 1995}
\def\Hunt85{Hunt, 1985}
\def\Nokes88{Nokes \& Wood, 1988}
\def\Bro97{Brown {\it et al.}, 1997}
\def\Hanna99{Hanna {\it et al.}, 1999}
\def\Olesen{Olesen, 1994}

\title{Characteristics of scalar dispersion in turbulent-channel flow}

\author{Juan C. del \'Alamo
  \footnote{School of Aeronautics UPM, 28040 Madrid, Spain.
    }
  \and
   Javier Jim\'enez
  \footnote{Also at School of Aeronautics UPM, 28040 Madrid, Spain.}
  }

\begin{document}

\maketitle
The dispersion of a passive scalar by wall turbulence, in the limit of infinite
Pecl\'et number, is analyzed using frozen velocity fields from the DNS by
\cite{alajim01}. The Lagrangian trajectories of fluid particles in those fields
are integrated and used to compute the first and second-order moments
of the distribution of fluid-particle displacements.
It is shown that the largest scales in the flow dominate turbulent diffusion,
and the computed dispersions are in good agreement with measurements in the
atmospheric boundary layer. This agreement can be understood noting that
the life times of the large strucutures are much longer
than the time scale of the transition from linear to Gaussian particle
spreading in the cross-stream plane. 
Numerical experiments performed computing the Lagrangian trajectories in
reference frames moving at different velocities suggest that this transition
is controlled by the difference between the mean streamwise velocity and the phase
speed of the large-scale structures of the cross-stream velocities. 
In the streamwise direction, the effect of the mean shear dominates and
produces elongated scalar patches, with dispersion exponents which are different
from the transverse ones.

\vskip0.1in
\hrule
\section{Introduction}
The prediction of the diffusion characteristics in turbulent shear flows,
particularly in those near walls, is a notoriously difficult problem. While
for example, the width of a contaminant plume 
follows relatively well a Gaussian spreading law in isotropic turbulence,
or even in wall-bounded flows when measured far enough from the
source, the same is not true when the spreading is measured closer to the
source (\Nokes88), near the wall, or in atmospheric flows.
This is an important consideration in many practical applications, such as in
the prediction of dispersal of pollution from industrial plants, or of hazardous
substances from either accidental or malicious releases. There are many other
problems in which this subject is important, apart from the ones already
mentioned. For instance, the diffusion of odors in the atmosphere is known to
affect the migrational patterns of some insects, and it is not known whether
similar effects occur in other anisotropic flows, such as near-surface ocean
turbulence, where it could influence the rate of decay of the thermal 
wake of vehicles. The solution to 
these problems is typically estimated using empirical laws (\Bro97), 
or computed from semi-empirical models (\Hanna99). Many of these models
are used for regulatory purposes, and
the fact that some of them produce different results for the same
input data is an indication of the difficulty of the problem. This has
led to the development of standardization programs (\Harmoni) with the
purpose of establishing systematic procedures for the development and testing 
of dispersion models, based on compilations of meteorological data from
field experiments. However, due to the inherent difficulty of performing 
such experiments, the data sets are scarce, the number of measured
magnitudes is limited and some of them are of doubtful accuracy (\Olesen). 

Since the atmospheric effects are observed over scales of hundreds of meters, 
and there are sound theoretical arguments to expect small-scale turbulence
to produce Gaussian diffusion at such long distances, it is tempting to 
conclude that the reason for the anomalous spreading is the presence of the
very large anisotropic scales (VLAS) in turbulent wall flows. Recently we have
performed a direct numerical simulation of turbulent channel flow at moderate
Reynolds number, which we believe to be the first one in which both the Reynolds
number is high enough to observe some scale separation, and in which the 
numerical box is large enough not to interfere with the dynamics of the largest
scales. The present work, which used flow data from this simulation, is intended
to be a first step in using direct numerical simulation in the subject of
atmospheric dispersal, which might contribute to diminishing the current
experimental uncertainties.
\begin{table}
\begin{center}
\begin{tabular}{ccccc}
\quad Case \quad & \quad $U_{adv}/U_b$ \quad & \quad  Spatial Resolution \quad &
\quad No. of Fields \quad & \quad No. of Particles per Field \quad \\ 
\hline
\hfill
$1$ & $0$ & full & $1$ & $2\times 10^5$ \\
\hfill
$2$ & $0.84$ & full & $1$ & $2\times 10^5$ \\
\hfill
$3$ & $0$ & $\lambda_x,\,\lambda_z > 0.25\, h$ & $3$ & $2\times 10^5$ \\
\hfill
$4$ & $0.84$ & $\lambda_x,\,\lambda_z > 0.25\, h$ & $3$ & $2\times 10^5$ \\
\hfill
\end{tabular}
\caption{Summary of computed cases.}
\label{tab:tabla}
\end{center}
\end{table}
\section{Computing dispersion from frozen fields}
\label{sec:frofi}
We will consider the release of a passive scalar into turbulent-channel flow
in the limit of infinite Pecl\'et number $U_b h /D$ (here $U_b$ is the bulk
mean velocity in the channel, $h$ is the channel half-width, and $D$ is the
kinematic diffusivity of the scalar). In that case the dispersion of the scalar
is controlled by the Lagrangian trajectories $\bf x$ of the fluid elements that
transport it, given by
\begin{equation}
\frac {\rm d {\bf x}} {\rm d t} = {\bf u}({\bf x}(t),t).
\label{eq:lagrantrue}
\end{equation}
The main difficulty of computing the Lagrangian trajectories of fluid particles 
lies in knowing the unsteady three-dimensional velocity field $\bf u( \bf x,t)$,
which has to be computed from the continuity and the Navier-Stokes equations,
leading to a problem much more expensive than the integration of
(\ref{eq:lagrantrue}) itself. Due to the preliminary nature of this work, and
in order to avoid the computational expense of integrating in time the
Lagrangian trajectories coupled with the velocity field, we have decided to
calculate the former using frozen velocity fields that were already available
from the DNS of turbulent-channel flow by \cite{alajim01}. This simulation was 
performed at a Reynolds number $Re_\tau=550$ based on the friction velocity
$u_\tau$ and on the channel half-width $h$, and its most important
characteristic is that the computational domain is large enough not to interfere
with the largest scales in the flow, which will allow us to study their effect
on the scalar dispersion. The size of the numerical box is
$L_x \times L_y \times L_z \,=\, 8\pi h \times 2h \times 4\pi h$ in the
streamwise, wall-normal and spanwise directions, respectively. In isotropic
turbulence the frozen-field approximation would be reasonable for times much
shorter than the characteristic life time of the eddies, which is proportional
to their turnover time. However, this might not be true in wall turbulence,
where the flow features are known to travel in the streamwise direction with an
advection velocity of the order of $U_b$ (\Wills). This advection velocity,
acting on scales of length $\lambda$, introduces a convective time scale
$T_c \sim \lambda / U_b$ that is always shorter that the eddy-turnover time
$T_L \sim \lambda / u_\tau$. We have tried to take into account the effect of
the mean advection by integrating the Lagrangian trajectories from the frozen
velocity fields in a moving reference frame,
\begin{figure}
  \begin{center}
    \includegraphics[width = 0.65\textwidth,
    angle = 0] 
    {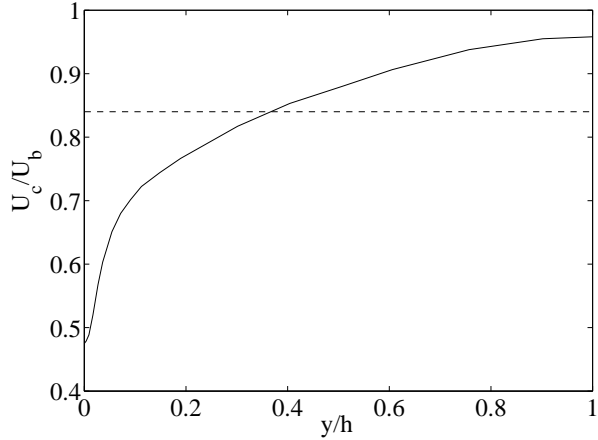}
    \caption{\solid, phase velocity $U_c$ of the large-scale spanwise velocity 
    component as a function of wall distance. Only structures such that 
    $\lambda_x, \lambda_z\ge 0.25\,h$ are taken into account. 
    \dashed, advection velocity $U_{adv}$ of the frozen fields.}
    \label{fig:uphase}
  \end{center}
\end{figure}
\begin{equation}
\frac {d {\bf x}} {d\tau} = {\bf u}\left[{
{\bf x (\tau)-{\bf \tau U_{adv}}},\,t_0}\right].
\label{eq:lagranapprox}
\end{equation}
Here $\bf u({\bf x},t_0)$ is the instantaneous frozen velocity field at $t=t_0$,
and ${\bf U_{adv}} = (U_{adv},0,0)$ is the velocity of the reference frame,
which can be interpreted physically as a choice for the advection velocity
of the frozen fields. This choice affects the paths of fluid particles
by modifying their velocities relative to the turbulent structures.
In order to evaluate the effect of the advection velocity of the frozen
fields in scalar dispersion, we have integrated (\ref{eq:lagranapprox})
for two different values of $U_{adv}$. In one case we have chosen $U_{adv}=0$,
while in the other one we have set it to be equal to the representative phase
velocity of the large energetic scales in the flow, which have widths and
lengths of the order of or larger than $h$ (\alajim01). There are several
possible ways to compute the phase velocity of a flow variable (\Wills; \HUSS;
\AJ02). Here we have computed it from the frequency-wave number power spectrum
$P(\omega,k_x,y)$ as in \cite{Wills64}, where the phase velocity is defined as
the velocity $U_c(y)$ of the moving reference frame for which the integral time
scale
\begin{equation}
T_L(k_x,y) = \frac {P(-U_c k_x, k_x,y)} { \int_{-\infty}^{\infty}P(\omega,k_x,y)
\rm d \omega }
\label{eq:lagrantime}
\end{equation}
is maximum. The frequency-wave number power spectrum has been computed using
time histories of velocity fields that were available from the DNS, similarly
to \cite{ChoiMoin90}. In the present case, the maximum and the minimum
frequencies imposed by the temporal sampling are $\omega_{min}=0.14\, U_b/h$ and
$\omega_{max}=71 \, U_b/h$. Due to storage limitations, the time histories of
the velocity field were spatially filtered by removing all the length scales
either shorter or narrower than $0.25\,h$ using a Fourier cut-off filter.
Figure \ref{fig:uphase} displays the average phase speed of the low-pass
filtered spanwise velocity fluctuations as a function of wall-distance, and 
non-dimensionalized with the bulk mean velocity $U_b$ (solid line), together
with its average across the channel width (dashed line)
$$ U_{adv} = \frac 1 {2h} \int_0^{2h} U_c(y) \rm dy = 0.84 \, U_b.$$
We have chosen this value as the advection velocity of the frozen fields to be
used in (\ref{eq:lagranapprox}) for our second set of experiments.
\cite{HussKim93} computed the propagation speeds of several turbulent 
magnitudes, including the velocity components, in a fully-resolved $Re_\tau=180$
channel. They obtained a phase velocity of $w$ in the near-wall region 
approximately equal to $10\,u_\tau$, which is the same that we have measured in
the low-pass filtered $Re_\tau=550$ channel. In the outer region, however,
they obtain advection velocities approximately $10\%$ higher than we do, and 
which are closer to the local mean velocity in their case than in ours. This is
not surprising. If we think that turbulent structures propagate roughly
at the average streamwise velocity that they feel, then the smaller
scales should follow the local mean velocity better than the large ones.

\begin{figure}
  \begin{center}
    \includegraphics[width = 0.65\textwidth,
    angle = 0] 
    {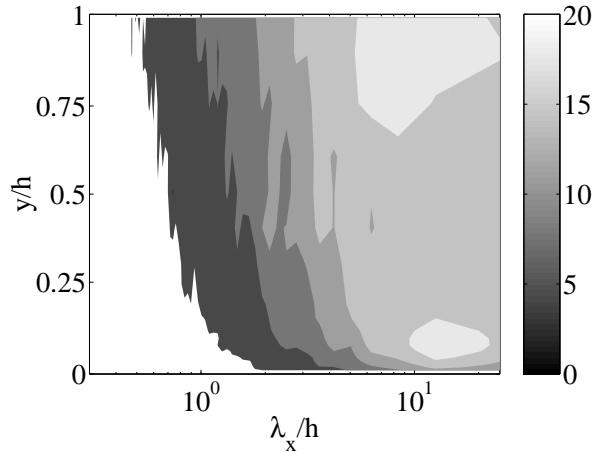}
    \caption{Lagrangian time scale $U_b T_L / h$ of the spanwise velocity,
             low-pass filtered in $z$ ($\lambda_z>0.25\,h$),
	     as a function of the streamwise wavelength $\lambda_x$
             and wall distance $y$.}
    \label{fig:tlagran}
  \end{center}
\end{figure}

The integral time scale $T_L$ in (\ref{eq:lagrantime}) measures the 
characteristic time associated to the turbulent fluctuations of a given 
magnitude with a certain length $\lambda_x = 2 \pi / k_x$ at a given 
wall-distance, and in a reference frame moving with their local advection
velocity. This magnitude can be interpreted as the Lagrangian time scale
seen by an observer following the mean trajectories of the eddies, or in other
words, as the typical lifetime of the structures of a given length. The
Lagrangian time scale of the fluctuations of spanwise velocity has been
represented in figure \ref{fig:tlagran}, low-pass filtered in $z$, as a 
function of streamwise wavelength $\lambda_x$ and wall distance $y$. The figure
shows that the lifetimes of the large scales of $w$ can be very long, even
comparable to a wash-out time $8\pi\, h / U_b$. The values of $T_L$ for the
other two components of velocity, not shown here, are similar, and give an
{\it a-priori} estimate of the longest intervals of time for which we can 
expect the  integration of (\ref{eq:lagranapprox}) to provide reasonably
accurate results. Note that this prediction would only be true if the large
scales controlled the characteristics of dispersion. In order to analyze their
importance in this phenomenon, we have solved (\ref{eq:lagranapprox}) using
both fully-resolved and cut-off filtered fields. Overall, we have integrated
(\ref{eq:lagranapprox}) in four different cases, depending on the choice of
$U_{adv}$ and of the spatial resolution. These cases have been summarized in
table \ref{tab:tabla}, indicating the number of different fields that have been
used for each case, as well as the number of trajectories that have been
computed per field.
The time discretization is fourth-order Runge-Kutta, and third-order B-splines
have been used to interpolate the velocity field from the collocation points of
the DNS.

\section{Results. One-point statistics}
\begin{figure}
  \begin{minipage}[t]{0.49\textwidth}
    \includegraphics[width = \textwidth]{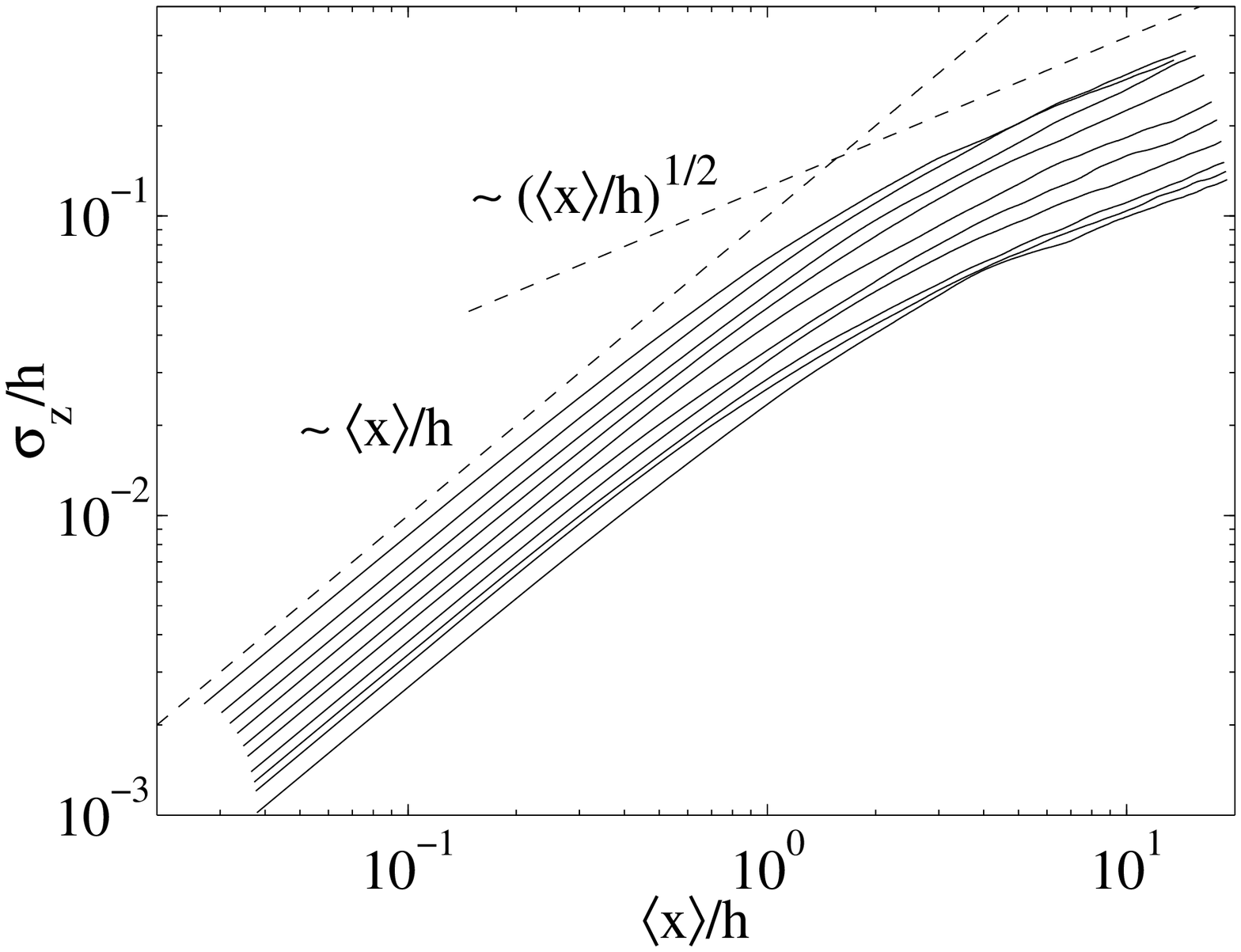}    
    \mylab{.2\textwidth}{.7\textwidth}{\it (a)}%
  \end{minipage}
  \hfill
  \begin{minipage}[t]{0.49\textwidth}
    \includegraphics[width = \textwidth]{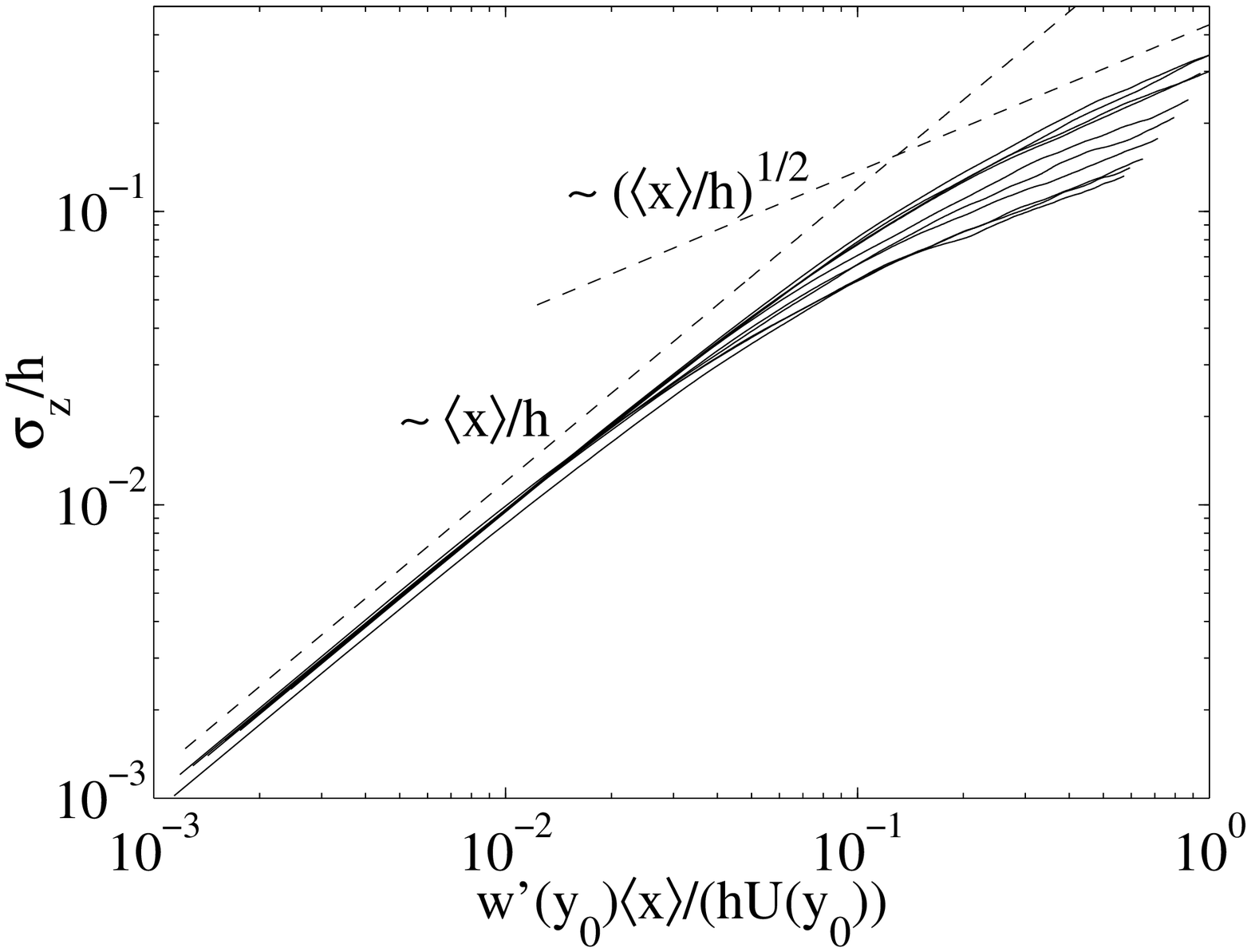}    
    \mylab{.2\textwidth}{.7\textwidth}{\it (b)}%
  \end{minipage}
  \caption{R.m.s. of the spanwise displacement $\sigma_z$ of fluid 
           particles as a function of their mean streamwise displacement 
	   $\langle x \rangle$. The different curves correspond to 
           $y_0/h=0.1(0.1)1$ from bottom to top. The dashed lines have 
           logarithmic slopes $1$ and $1/2$. 
	   {\it (a)}, using $h$ as the scale of $\sigma_z$ and $\langle x 
           \rangle$;
           {\it (b)}, using $w'(y_0)t$ as the scale of $\sigma_z$ and
	   $U(y_0)t$ as the scale of $\langle x \rangle$. 
	   }
  \label{fig:sigmacollapse}
\end{figure}
The single-particle statistics $\langle x_i \rangle$ and 
$\sigma_i = \langle (x_i - \langle x_i \rangle)^2 \rangle^{1/2}$ 
are of great interest because they indicate respectively the mean displacement 
of the center of a typical scalar patch and its size in the three directions
of space, and also because the latter is often measured as a function of
the former in field experiments, which will allow us to test the approximation
(\ref{eq:lagranapprox}). These magnitudes are functions of the initial position
$y_0$ of the fluid element, of its instantaneous position $y$, and of time.
Operating on (\ref{eq:lagrantrue}) it is possible to obtain 
(\Hunt85) that
\begin{equation}
\partial_t \sigma_i^2 = u_i'(y)u_i'(y_0) 
\int_0^t \rho_{ii}(r_x-\tau U_{adv},r_z,y_0,y,t-\tau) \rm d \tau,
\label{eq:correldisp}
\end{equation}
where $\rho_{ii}$ is the two-point autocorrelation coefficient of the
{\it i}\/th-component of the velocity vector. This magnitude is a function of
the streamwise and spanwise separations $r_x$ and $r_z$, of the initial
and instantaneous wall-distances, and of time.
Note that the frozen-field approximation is equivalent to 
setting $t-\tau=0$ in $\rho_{ii}$ in (\ref{eq:correldisp}). 
For times and spatial separations short compared to the corresponding integral
scales, the velocity field is almost fully correlated, $\rho_{ii} \approx 1$,
and $y_0 \approx y$. We then have
\begin{equation}
\sigma_i \approx u_i'(y_0) t \sim \langle x \rangle \approx U(y_0)t.
\label{eq:scaling}
\end{equation}
On the other hand, for very long temporal and spatial separations the velocity
field is approximately decorrelated, $\rho_{ii} \approx 0$, and the integral
in the right-hand side of (\ref{eq:correldisp}) is roughly independent of its
upper limit. We then obtain the Gaussian-spreading law 
$\sigma_i \sim (\langle x \rangle h)^{1/2}$.
Both asymptotic behaviors can be observed in figure \ref{fig:sigmacollapse},
where the solid curves show $\sigma_z$ from case 2 as a function of the
mean streamwise displacement for ten equispaced intervals of initial
wall-distances, from the wall to the center of the channel. In figure 
\ref{fig:sigmacollapse}\/{\it (a)} we have used the channel half-width as the
length scale for $\sigma_z$ and $\langle x \rangle$, while in figure 
\ref{fig:sigmacollapse}\/{\it (b)} we have scaled $\sigma_z$ with $w'(y_0)t$
and $\langle x \rangle$ with $U(y_0)t$. The figures show that in the
short-range limit, the curves representing $\sigma_z$ are
parallel to the dashed line with logarithmic slope $1$, while far away from the
release point the curves are roughly parallel to the dashed line with
logarithmic slope $1/2$. It can be observed in figure
\ref{fig:sigmacollapse}\/{\it (b)} that the scaling (\ref{eq:scaling}) collapses
well the plume width corresponding to different release points, at least at 
short distances from the source. As expected, the collapse worsens
beyond the turning point in the curves, where their slope starts decreasing and 
(\ref{eq:scaling}) is no longer valid. The characteristic position of this
turning point is a measure of the shortest integral scale involved in the
dispersion process.
\begin{figure}
  \begin{center}
    \includegraphics[width = 0.65\textwidth,angle = 0] {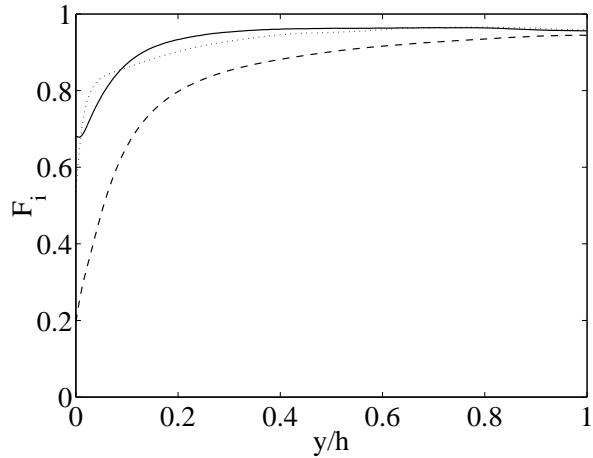}
    \caption{Fraction $F_i$ of the {\it i}\/th-component of turbulent kinetic
    energy contained in the cut-off filtered fields as a function of 
    wall-distance. \solid, $u$; \dashed, $v$; \dotted, $w$.}
    \label{fig:fracener}
  \end{center}
\end{figure}

\begin{figure}
  \begin{minipage}[t]{0.49\textwidth}
     \includegraphics[width =\textwidth, angle=0]{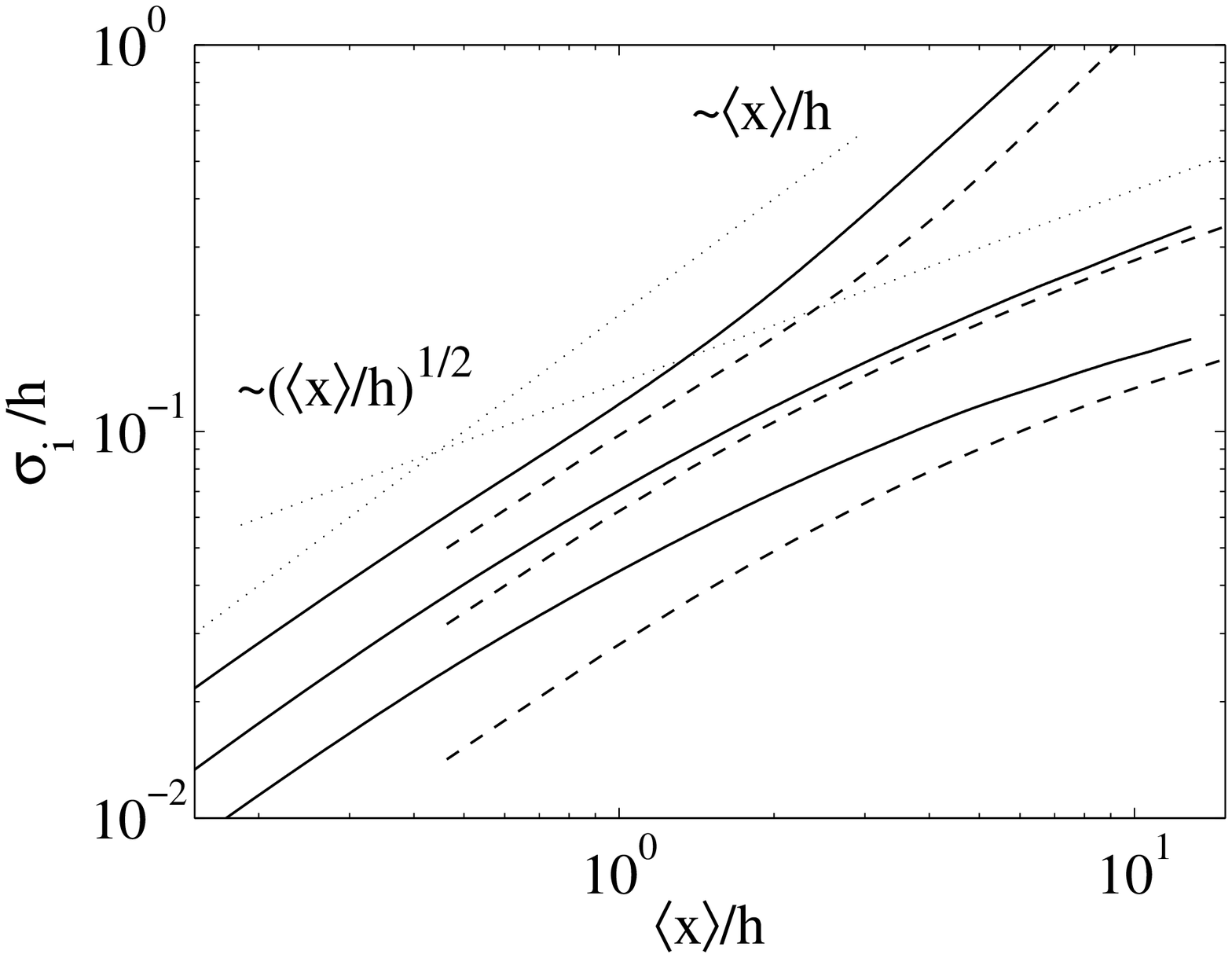}
     \mylab{.2\textwidth}{.7\textwidth}{\it (a)}%
  \end{minipage}
  \begin{minipage}[t]{0.49\textwidth}
     \includegraphics[width =\textwidth, angle=0]{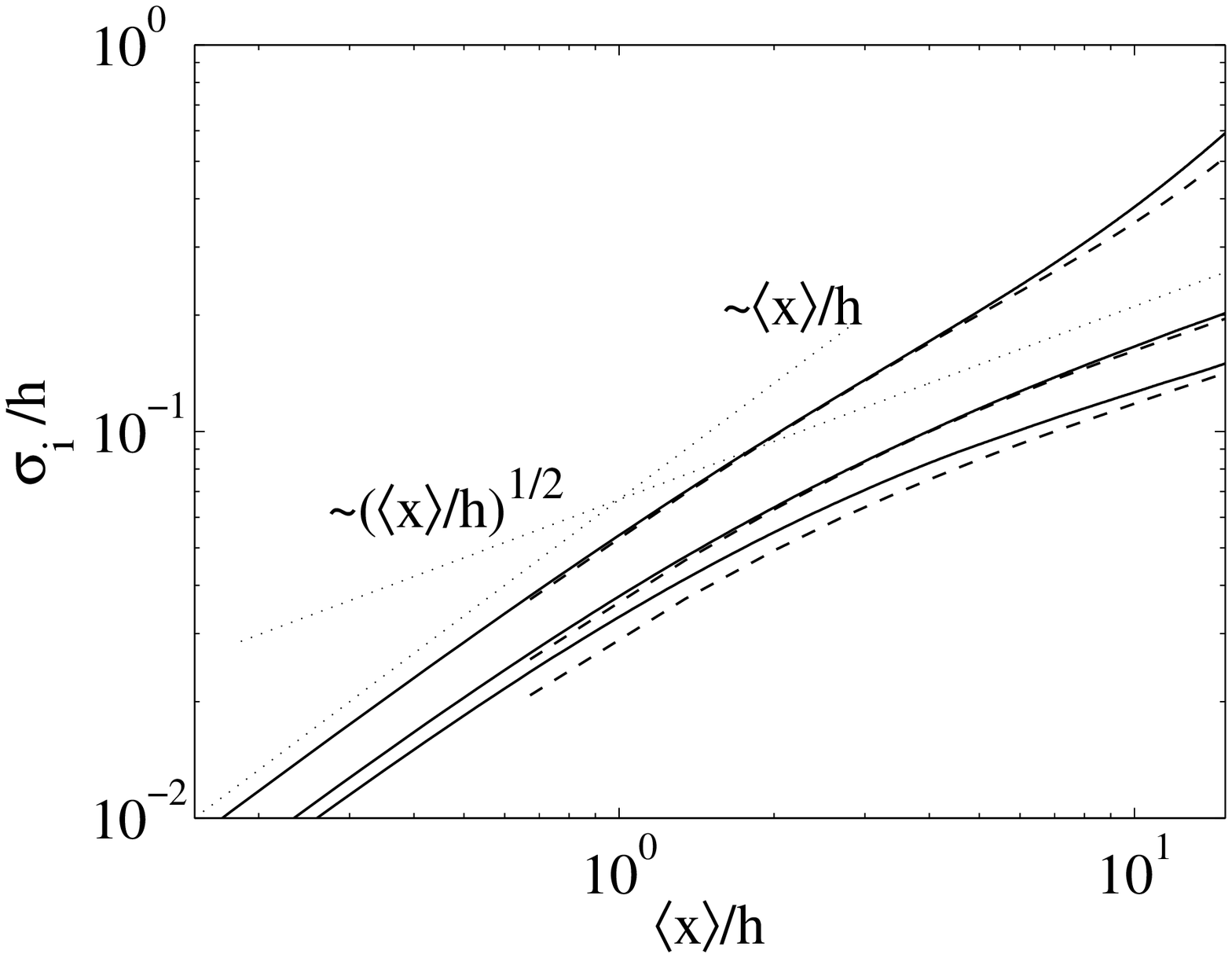}
     \mylab{.2\textwidth}{.7\textwidth}{\it (b)}%
  \end{minipage}
  \caption{R.m.s. of the displacement of fluid particles in the
          {\it i}\/th-direction as a function of their mean streamwise
          displacement $\langle x \rangle$. Top, $\sigma_x$; middle, $\sigma_z$;
          bottom, $\sigma_y$. \solid, full DNS fields (case 2); \dashed, cut-off
          filtered DNS fields (case 4). In all the cases $U_{adv}=0.84\,U_b$.
          {\it (a)}, $y_0^+<100$; {\it (b)}, $0.2 <y_0/h<1$.}
  \label{fig:sigmacompraw}
\end{figure}

Equation (\ref{eq:correldisp}) also suggests that the large scales may play an
important role in turbulent dispersion. Coherent structures with 
$\lambda_x/h > 2$ and $\lambda_z/h \approx 1-2$ are known to be correlated
all across the channel half-width and to contain a large fraction 
of the turbulent kinetic energy (\AJ02), which suggests that
they should contribute substantially to the right-hand side of 
(\ref{eq:correldisp}). Figure \ref{fig:fracener} displays the fraction $F$ of
the total streamwise (solid line), wall-normal (dashed line) and spanwise
(dotted line) kinetic energy contained in the cut-off filtered fields as a
function of wall-distance. The figure shows that the structures which are longer
and wider than $0.25\,h$, contain most of the kinetic energy of $u$ and $w$ in
the outer region of the flow, and hence could be expected to produce values
of $\sigma_x$ and $\sigma_z$ similar to the ones generated by the full fields.
On the other hand, the small scales of $v$ contain relatively more kinetic
energy than those of $u$ and $w$, suggesting that the value of $\sigma_y$ 
computed from the filtered fields should be less approximate to the one obtained
from the full fields. This is actually what is observed in figure
\ref{fig:sigmacompraw}, where we have plotted the three components of $\sigma$
(from top to bottom $\sigma_x$, $\sigma_z$ and $\sigma_y$) computed from the
full (case 2, solid lines) and the filtered (case 4, dashed lines) moving
frozen DNS fields. In figure \ref{fig:sigmacompraw}\/{\it (a)} the patch size
has been averaged for fluid particles released in the near-wall region
($y_0^+ < 100$), while in figure \ref{fig:sigmacompraw}\/{\it (b)} the average
has been performed for initial positions in the outer region ($0.2< y_0/h<1$).
The results from the filtered fields compare fairly well to those from the
fully resolved ones in  the outer region, while they underestimate the different
components of the r.m.s. in the near-wall region. Note that the agreement
between the different sets of data is better wherever $F$ is higher and 
{\it vice-versa}, supporting the argument above. These observations agree with
the previous work by \cite{Arme99}, who performed a similar analysis using
time-evolving velocity fields, with application to LES modeling. The results
from the stationary frozen DNS fields (cases 1 and 3), not shown here, behave
in the same way as the ones we have presented in figure \ref{fig:sigmacompraw}.

Note that the standard deviations in figure \ref{fig:sigmacompraw} are always
much lower than $\langle x \rangle$, implying that the basic motion of
the particles is advection by the local mean velocity, 
$\langle r_x \rangle \approx U(y_0)\tau$, while the spreading around that 
position is slow.

\begin{figure}
  \begin{minipage}[t]{0.49\textwidth}
    \includegraphics[width = \textwidth]{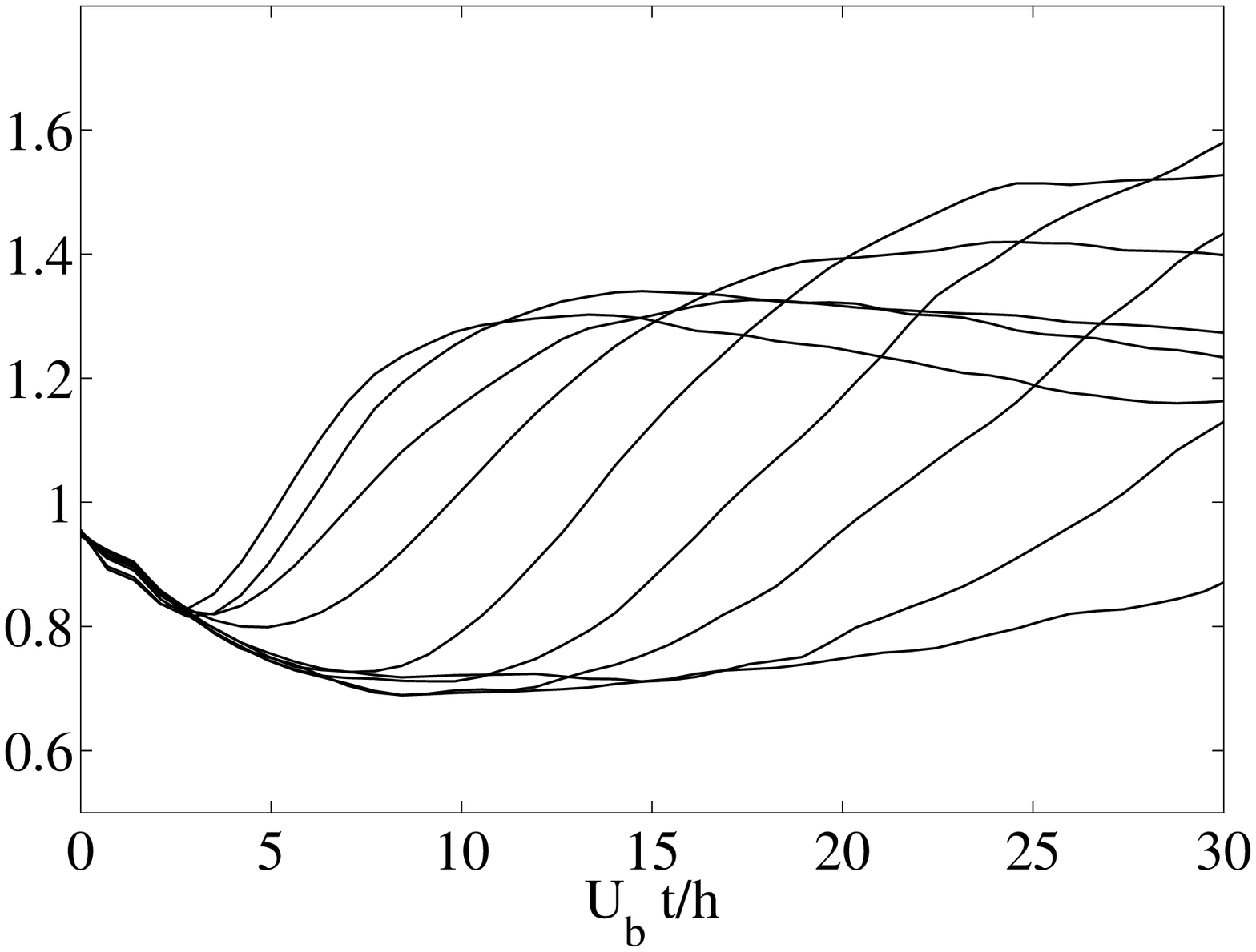}    
    \mylab{.1\textwidth}{.7\textwidth}{\it (a)}%
  \end{minipage}
  \hfill
  \begin{minipage}[t]{0.49\textwidth}
    \includegraphics[width = \textwidth]{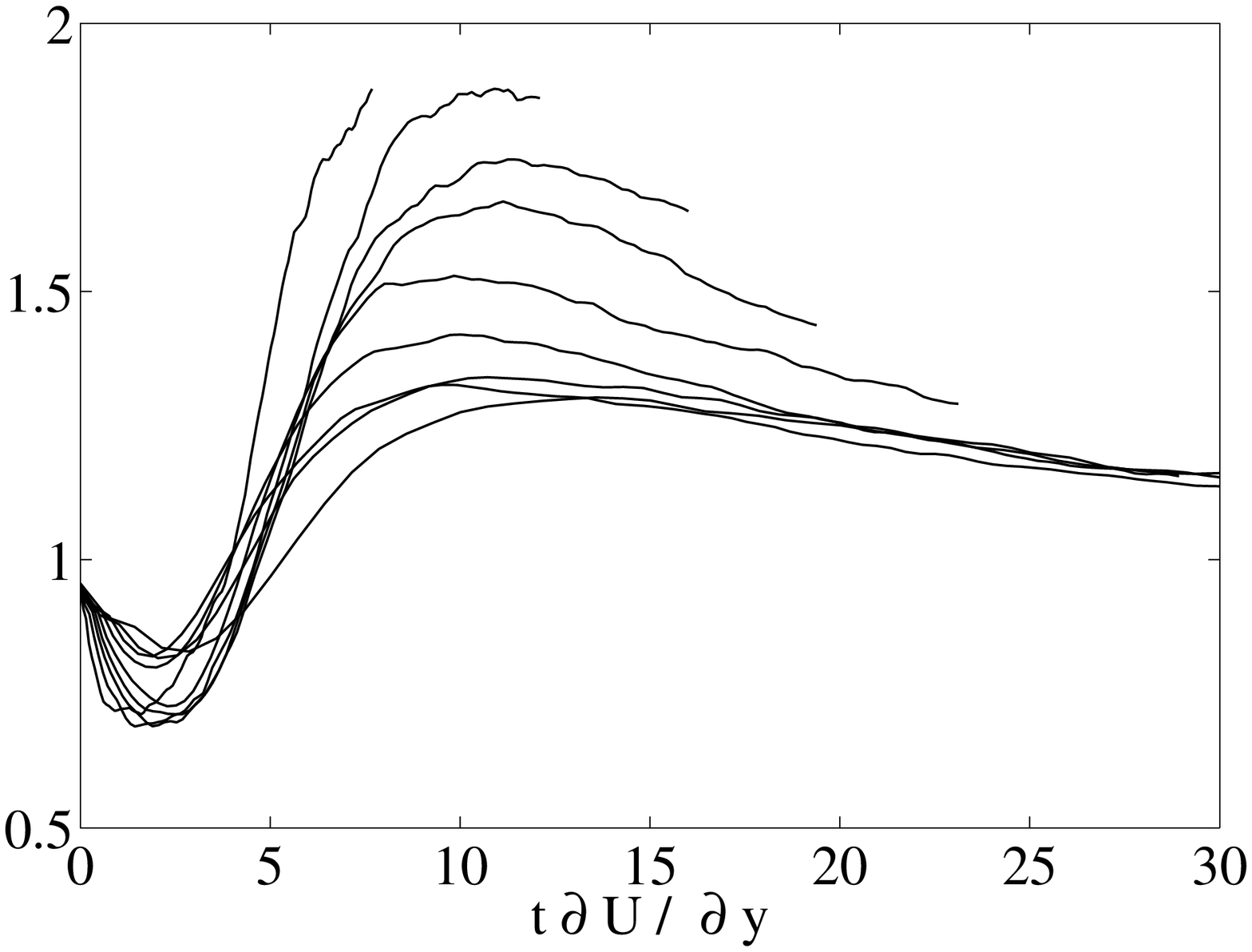}    
    \mylab{.1\textwidth}{.7\textwidth}{\it (b)}%
  \end{minipage}
  \caption{Logarithmic slope of $\sigma_x$ (case 2) as a function of time $t$
           for $y_0/h=0.1(0.1)1$. {\it (a)}, using time non-dimensionalized 
           with $U_b/h$. The curves peak at longer times as we move away from
	   the wall; {\it (b)}, using time non-dimensionalized 
           with $\partial_y U$. The values of the maxima increase as we move
	   away from the wall.}
  \label{fig:exponentesx}
\end{figure}

Figure \ref{fig:sigmacompraw} also shows the relative magnitude of $\sigma$
along the different axes, which can help us have an idea about the
evolution of the shape of a typical scalar patch with the distance to the 
source. In the short-range region the three standard deviations grow at
the same rate, and a typical cloud of scalar would initially conserve its
original shape as it moves away from the release point. However, after the
cloud has traveled a certain distance it would start elongating very rapidly,
as we can deduce from the increase in the slope of $\sigma_x$ that takes place
in the figure at long distances. The comparison of figures
\ref{fig:sigmacompraw}\/{\it (a)} and \ref{fig:sigmacompraw}\/{\it (b)}
indicates that this phenomenon occurs at a shorter distance to the source for
lower values of $y_0$, which is more apparent in figure \ref{fig:exponentesx}.
This figure displays the logarithmic slope of $\sigma_x$ (from case 2) as a
function of time for ten  equispaced intervals of $y_0$. The slope of 
$\sigma_x$ increases and reaches a maximum value in times which are longer as
the curves move from the left to the right, which correspond to increasing
values of $y_0$. There are strong arguments to think that this effect
is due to the mean shear. Scaling time with $\partial_{y} U$ in
figure \ref{fig:exponentesx}\/{\it (b)} makes the position of the maxima 
of the different curves collapse, indicating that the time scale associated to 
this phenomenon is the inverse of the mean shear. Besides, in a different
experiment we integrated (\ref{eq:lagranapprox}) using fields from which we had
removed the mean velocity profile, and the resulting $\sigma_x$ behaved in the
same way as $\sigma_y$ and $\sigma_z$, and did not show the transient increase
in slope. \cite{Tennekes} show that in a flow subjected to a uniform shear $S$,
the  dispersion in the streamwise direction increases asymptotically with time
as $(St)^{3/2}$. This value of the logarithmic slope lies roughly in the center
of the set of different maximum values that we obtained from the DNS fields,
and the scatter in the numerical values might be explained by the fact that
$\partial_y U$ is not uniform in a turbulent channel.

\subsection{Comparison with atmospheric data}

\begin{figure}
  \begin{center}
    \includegraphics[width = 0.75\textwidth,
    angle = 0] 
    {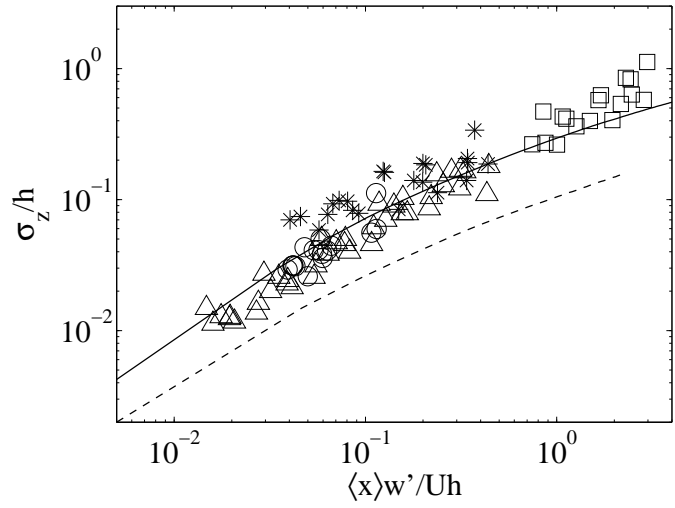}
    \caption{R.m.s. of the spanwise displacement of fluid particles,
             $\sigma_z /h$, as a function of their mean streamwise displacement
             $\langle x \rangle w' / ( U h)$. Lines, data from DNS fields;
             \solid, $U_{adv}=0.84\,U_b$; \dashed, $U_{adv}=0$. Symbols, 
             atmospheric experiments; \circle, Ecofin Project (\Risoe);
             \trian, Prairie Grass (\Harmoni);  $*$, 
             Lillestr\o m; \squar, Copenhagen (\Harmoni). }
    \label{fig:compexp}
  \end{center}
\end{figure}

\begin{figure}
  \begin{minipage}[t]{0.49\textwidth}
    \includegraphics[width = \textwidth]{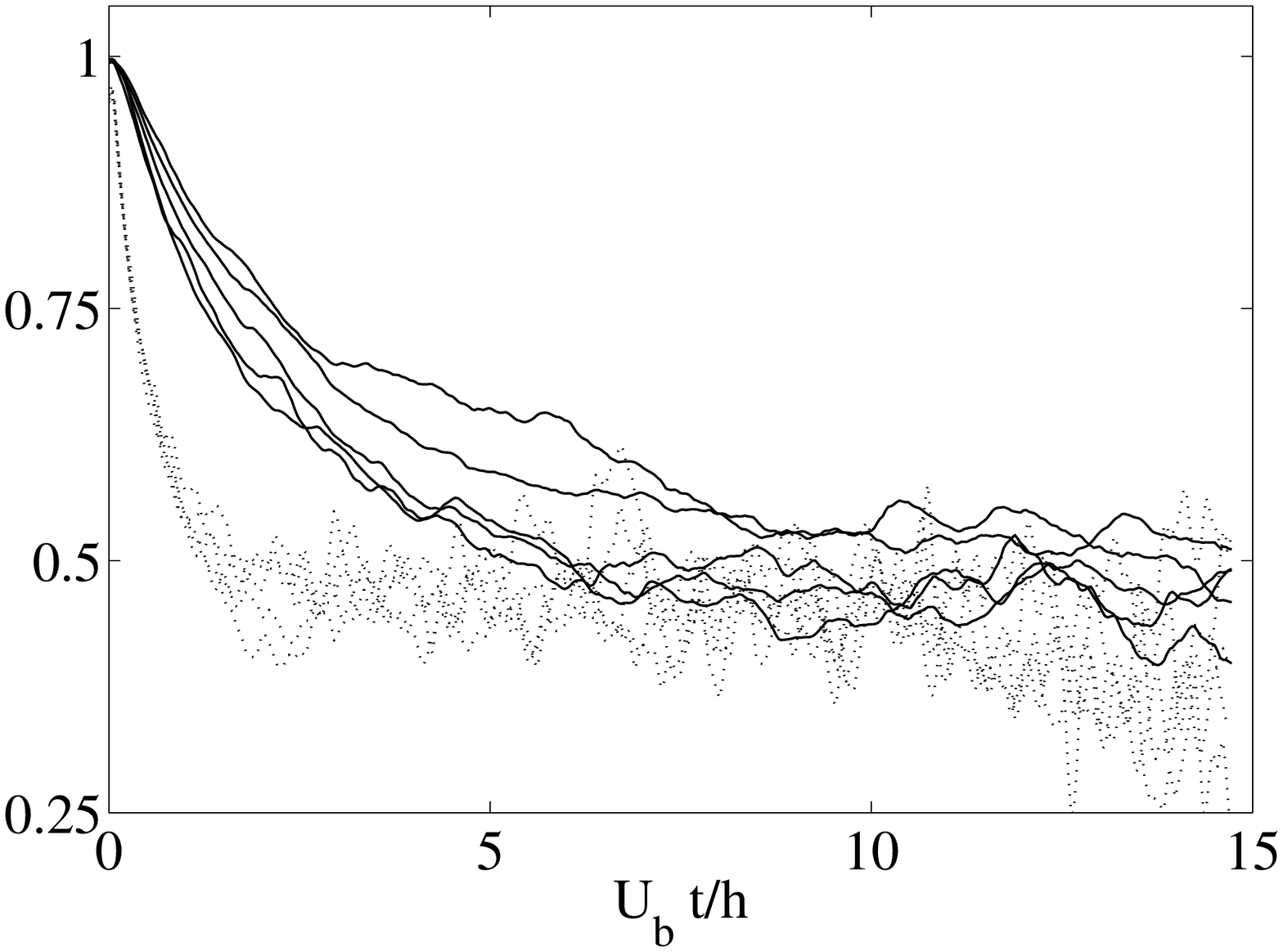}    
    \mylab{.8\textwidth}{.7\textwidth}{\it (a)}%
  \end{minipage}
  \hfill
  \begin{minipage}[t]{0.49\textwidth}
    \includegraphics[width = \textwidth]{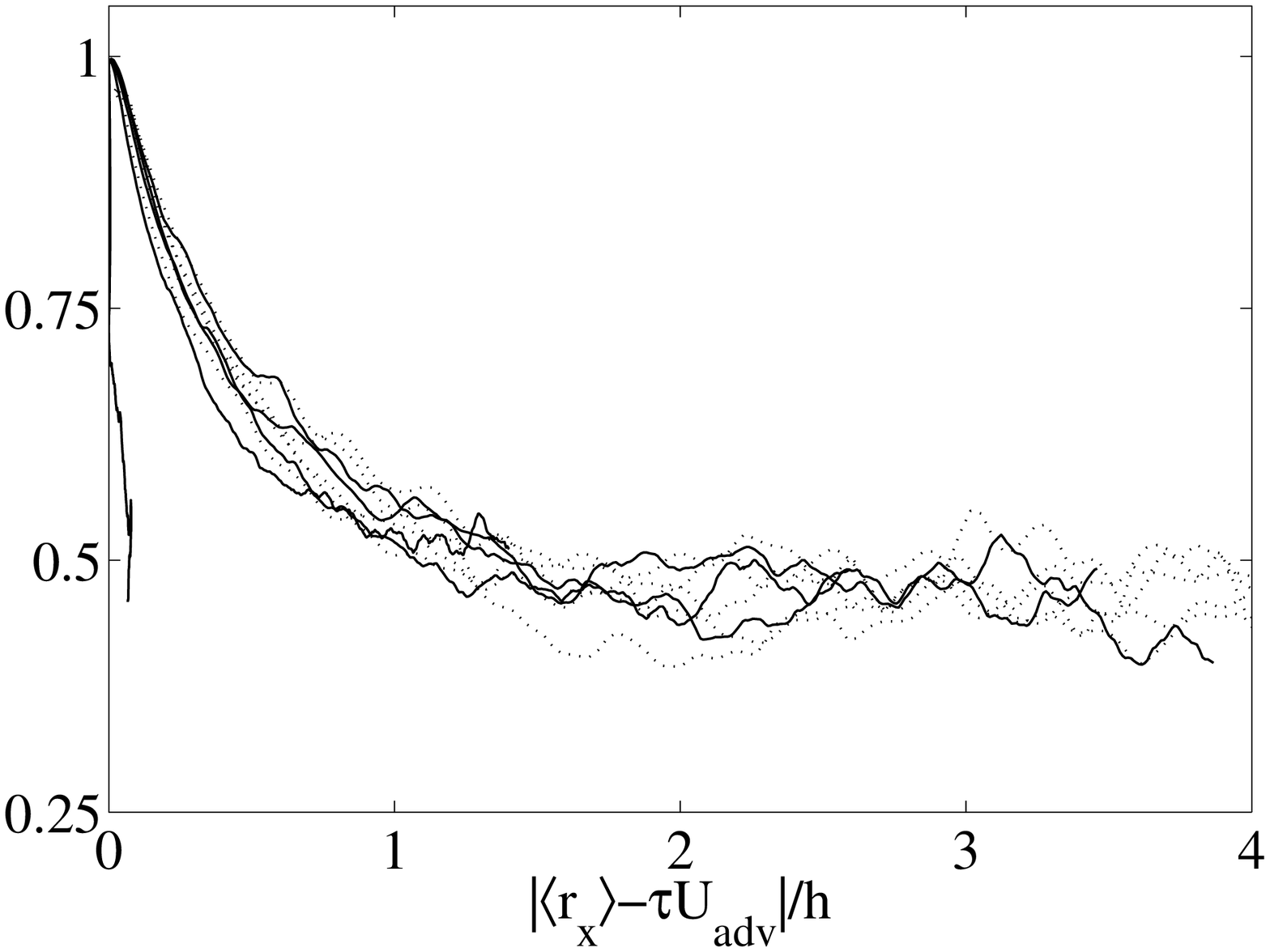}    
    \mylab{.8\textwidth}{.7\textwidth}{\it (b)}%
  \end{minipage}
  \caption{Logarithmic slope of $\sigma_z$ for $y_0/h=0.2(0.2)1$. 
           {\it (a)}, as a function of time $U_b t/h$. 
           {\it (b)}, as a function of the 
	   separation $\left| { \langle r_x \rangle - \tau U_{adv}}\right| /h$.
	   \solid, $U_{adv}=0.84\,U_b$. \dashed, $U_{adv}=0$. 
	   The solid line that does not collapse well
	   in the left-hand side of {\it (b)}
	   corresponds to $y_0=0.2\,h$, near the critical layer at 
           $y_c=0.23\, h$;
	   }
  \label{fig:exponentesz}
\end{figure}

In section \ref{sec:frofi} we discussed the {\it a-priori} validity of our
study, obtaining a rough estimate of the longest time intervals for which we
could expect reasonable results from the model problem (\ref{eq:lagranapprox}). 
Here we analyze {\it a-posteriori} the frozen-turbulence
approximation, by comparing the computed dispersion characteristics 
with those measured in the atmospheric boundary layer. Figure \ref{fig:compexp} 
displays the r.m.s. of the spanwise displacement of fluid particles as a
function of their mean streamwise displacement. The symbols come from field
experiments, most of which were compiled by \cite{Niel02} and \cite{Oles95}.
The atmospheric data sets are difficult to compare among themselves and to
the numerical results. In general, the experiments consist of releasing a
passive tracer from a smokestack and measuring its near-ground concentration
along arcs situated at increasing distances from the release point. However,
neither the releases nor the measurements were performed at the same ground
distances in the different experiments, the monitoring procedures also differed,
and so did the topological and meteorological conditions. Thus, any quantitative
conclusion from the observation of figure \ref{fig:compexp} should be taken as
orientative, as it is also suggested by the dispersion of the data in the
figure.
The solid line comes from our numerical results with $U_{adv}=0.84 \, U_b$,
while the dashed line corresponds to $U_{adv}=0$. In both cases we have
represented the average values over the interval of particle positions
$y/h < 0.1$, in order to compare with the atmospheric near-ground measurements.
It should be noted however, that the numerical results contain the 
contributions from particles released at all the possible wall distances across
the channel, while the particles were released at a single ground-distance in
the experiments. Even so, the agreement between the numerical and the
experimental results is reasonably good for the results from the moving fields. They underestimate somewhat $\sigma_z$ at long streamwise distances, but this
could be due to the fact that the atmospheric data in the corresponding
experiment (the Copenhagen data set, represented with squares) were taken under
unstably  stratified atmospheric conditions, with Monin-Obukhov lengths of the
order of $-100\, \rm m$ (\Olesen). The results from the stationary fields
look qualitatively correct, but they predict widths lower than the experimental 
values. This can be understood considering that, since the particles move
approximately with the mean flow velocity, the mean streamwise separation that
goes into the correlation function in (\ref{eq:correldisp}) is
$\langle r_x \rangle \approx U(y_0) \tau$, so that the first argument in the
autocorrelation coefficient of $w$ is 
$$
\langle {r_x} \rangle - \tau U_{adv} \approx (U(y_0)-U_{adv})\tau.
\label{eq:rxUad}
$$
In the stationary fields, fluid elements separate faster from their initial
positions with respect to the flow structures than in the moving ones, because
the difference between the mean velocity and that of the reference frame is
higher in the former than in the latter. This is true all across the channel,
except in the near wall region where $U(y_0)$ is small. The fluid particles
in the stationary snapshots therefore feel a more uncorrelated velocity field
than in the advecting cases, leading to lower values of $\sigma_z$. This
argument is supported by figure \ref{fig:exponentesz}, which shows the
logarithmic slope of $\sigma_z$ as a function of time (figure 
\ref{fig:exponentesz}\/{\it a}) and as a function of
$\left|{\langle {r_x} \rangle - \tau U_{adv}}\right|$ 
(figure \ref{fig:exponentesz}\/{\it b}), for five equispaced intervals of $y_0$
from the wall to the center of the channel. In figure
\ref{fig:exponentesz}\/{\it (a)} the logarithmic slope of $\sigma_z$ decreases
faster with time for the stationary fields (case 1), plotted with  dotted lines,
than for the moving ones (case 2), represented with solid lines. On the other
hand, the curves in figure \ref{fig:exponentesz} \/{\it (b)} collapse fairly
well except for the curve on the left-hand side of the plot, which corresponds
to $y_0/h = 0.2$. This position is near the critical layer at $y_c=0.23\,h$,
where $U(y_c) = U_{adv}$. In that region, the mechanism that we have just
described is weaker, and it is reasonable to think that the dispersion is
dominated by the turbulent velocity fluctuations and not by the mean advection. 
Note that in the case of the stationary fields the critical layer is located at
the walls, which may explain why all the dotted lines
collapse well in figure \ref{fig:exponentesz}\/{\it (b)}. The behavior that
we have observed in figure \ref{fig:exponentesz} is common to the results
from both the fully-resolved and the cut-off filtered fields.

As we have mentioned, the location of the point where the logarithmic
slopes of $\sigma_{y,z}$ decay from $1$ to their asymptotic value of $1/2$, is a
measure of the integral scale most relevant in turbulent dispersion in the
cross-stream plane.
The fact that the results from the moving frozen fields, which 
neglect the time evolution of turbulent structures, are able to predict 
the position of the turning region in the data from the field experiments,
can help us identify that integral scale. The large scales of $w$ have
lifetimes (see figure \ref{fig:tlagran}) which are approximately $4$ times
longer than the time scale of to the decay of the slope of $\sigma_z$
from the moving fields (figure \ref{fig:exponentesz}\/{\it a}). This suggests
that the temporal decay of the turbulent structures may not be important
in the decorrelation that the particles feel as they move in the flow. On the
other hand, the streamwise separation corresponding to the transition of
$\sigma_z$ shown in figure \ref{fig:exponentesz}\/{\it (b)}, is $r_x \approx h$.
This length is essentially equal to the position of the peak of the 
premultiplied energy spectrum of $w$ (\alajim01), which is a measure of its
streamwise integral scale. These observations also apply to the wall-normal
direction, for which the experimental information is much more scarce than in
the spanwise direction. They suggest that the transition in $\sigma_{y,z}$ may
be caused by the difference between the mean velocity of the flow, and the phase
speed of the velocity components in the cross-stream plane.
\section{Conclusions}
The results show that the large scales of turbulent channels play a very
important role in turbulent dispersion in the outer region of the flow, 
specially in the streamwise and spanwise directions. These structures contain a
large fraction of the turbulent kinetic energy, and they are correlated
across the full channel (\AJ02), so they are expected to contribute
substantially to the standard deviations in (\ref{eq:correldisp}).
Filtered velocity fields retaining only structures with 
$\lambda_x, \, \lambda_z > 0.25\,h$ produce more than $90\%$ of 
$\sigma_{x,z}$ and roughly $80\%$ of $\sigma_y$ in the outer region. These
results indicate that LES should be a valuable tool in the study of scalar
dispersion.

The transition from linear to Gaussian spreading is due to the decorrelation
of the velocity field along the Lagrangian trajectories of the particles.
The life times of the large scales of the spanwise velocity
are roughly $4$ times longer than the time scale of the decay in the slope of
the plume width from $1$ to $1/2$. Thus, the time evolution of turbulent
structures does not seem to be significant in the decorrelation process that
leads to Gaussian spreading, and to a first approximation, it may be possible
to study turbulent diffusion using frozen velocity fields. In fact, we have
integrated the Lagrangian trajectories of fluid particles from frozen velocity
fields, obtaining values of $\sigma_z$ that agree well with atmospheric
measurements. The agreement is better when the trajectories are computed in a 
reference frame moving with the average phase velocity of the large scales.
The stationary frozen fields, on the other hand, produce values of $\sigma_z$
lower than those from the field experiments. This is so because the 
decorrelation times experienced by the fluid elements in the stationary fields
are shorter than those in the moving ones. In both cases, the decay in the slope
of $\sigma_z$ takes place when the streamwise separation of the particles
relative to the velocity fields, $\langle r_x \rangle - \tau U_{adv}$, is 
roughly equal to the streamwise integral scale of $w$. Since the particles move 
in the $x$ direction following 
approximately the mean velocity profile, their separation with respect to their
initial positions is given by $(U(y_0)-U_{adv})\tau$, suggesting that the main
cause of the transition from linear to Gaussian spreading is the difference
between the mean streamwise velocity and the phase speeds of the velocity
components in the cross-stream plane.

The mean shear is the dominating mechanism in the streamwise direction. It
generates values of $\sigma_x$ much greater than the ones in the other two
directions, and leads to very elongated patch shapes. Although this
consideration is not important in the case of the dispersion of contaminants
from a continuous source, it may be fundamental in the case of discrete
releases. 

This work was supported by grant BFM 2000-1468 of CICYT. J.C.A. was supported by
the CTR and by the Spanish Ministry of Education. The authors would like to
thank Helge R. Olesen at NERI, Denmark for providing a digital version of the 
atmospheric data sets collected by the Harmonisation Group. We are also
indebted to Julian C. R. Hunt, with whom we had fruitful discussions related to
the subject of this work.

\end{document}